\documentclass[aps,twocolumn,showpacs]{revtex4}
\usepackage{graphicx}
\usepackage[all]{xy}
\usepackage{amsmath}
\usepackage{amssymb}
\newcommand{\be}{\begin{equation}}
\newcommand{\ee}{\end{equation}}
\newcommand{\bn}{\begin{eqnarray}}
\newcommand{\en}{\end{eqnarray}}
\newcommand{\bes}{\begin{subequations}}
\newcommand{\ees}{\end{subequations}}

\newcommand{\bb}{\bibitem}
\newcommand{\p}{\partial}
\begin{document}

\title{Two scalar field cosmology from coupled one-field models}
\author{P. H. R. S. Moraes$^{a}$ {{\footnote{moraes.phrs@gmail.com}}} and J. R. L. Santos$^{b}${{\footnote{dossantos.jrl@gmail.com}}}}
\affiliation{{\small {
$^a$INPE - Instituto Nacional de Pesquisas Espaciais - Divis\~ao de Astrof\'isica \\
Av. dos Astronautas 1758, S\~ao Jos\'e dos Campos, 12227-010 SP, Brazil \\
$^b$Departamento de F\'{\i}sica, Universidade Federal da Para\'{\i}ba, 58051-970 Jo\~ao Pessoa PB, Brazil\\
}
}}

\begin{abstract}

One possible description for the current accelerated expansion of the universe is quintessence dynamics. The basic idea of quintessence consists of analyzing cosmological scenarios driven by scalar fields. In this work we present some interesting features on the cosmological scenario obtained from the solutions of an effective two scalar field model in a flat space-time. This effective model was constructed by coupling two single scalar field systems in a nontrivial way via an extension method. The solutions related to the fields allowed us to compute analytical cosmological parameters. The behavior of these parameters are highlighted, as well as the different epochs obtained from them.
\end{abstract}

\pacs{98.80.Cq; 04.20.Jb}

\maketitle

\section{Introduction}
It is well known that presently the universe is undergoing a phase of accelerated expansion. The observational evidence for this phenomenon came from the study of Supernovae type Ia by the groups {\it Supernova Cosmology Project} \cite{riess/1998} and {\it High Redshift Supernova Team} \cite{perlmutter/1999}. In independent work, both groups expected supernovae brightness to be greater than their redshifts would theoretically suggest under the assumption of a non-accelerated expansion. However, they observed that the brightness was lower than predicted, unveiling an accelerated expanding universe.\

Since then many models have been proposed to explain theoretically this accelerating universe whose cause is named dark energy (DE). The most popular (and simplest) posits that the acceleration is due to quantum vacuum energy, described by the presence of a cosmological constant $\Lambda$ in the Einstein field equations. Although this model succeeds in explaining Supernovae Ia luminosity distance measures \cite{riess/1998,perlmutter/1999}, X-ray spectrum of cluster of galaxies \cite{allen/2004}, Baryon Acoustic Oscillations \cite{eisenstein/2005,percival/2010} and galaxy age data \cite{jimenez/2003}, when we compare the value of the quantum vacuum energy obtained from observational cosmology data \cite{hinshaw/2013} with the value computed using particle physics \cite{weinberg/1989}, the discrepancy between them obligates us to examine other DE models.\

Some useful sources in treating cosmic inflation \cite{guth/1981} and DE are the cosmological models involving scalar fields, which are the subject of this work. The theory of cosmological evolution based on scalar fields has been investigated in several areas covering the classical and the quantum level of the expanding universe. Moreover it has been considered in frameworks with inhomogeneous space-times, also known as stochastic inflation \cite{salopek}. This stochastic approach characterizes the quantum field fluctuations generation and evolution, which are supported by temperature fluctuations in the cosmic microwave background radiation. So far, many models describing the dynamics of the universe driven by a scalar field have been proposed \cite{brans/1961,mathiazhagan/1984,ratra/1988,peebles/1988,wetterich/1988,caldwell/1993,albrecht/2000,padmanabhan/2002} (and others) and in some cases, the scalar field $\phi(t)$ is named quintessence. The basic idea of quintessence consists of analyzing cosmological scenarios by adding a Lagrangian density of a scalar field (hereafter called ``Lagrangian") to the Einstein-Hilbert action. Some reviews on this subject can be found in \cite{fhsw,cds,sc,zws,alr,blrr} and references therein.\

By including a scalar field in the action, one obtains cosmological solutions based on the equations of motion, related to the dynamics of the field, which are second-order differential equations. Our proposal in this study is to work with analytical fields. We follow the investigation introduced in \cite{bglm}, where it was shown how to determine first-order differential equations involving one scalar field, whose solutions satisfy the equations of motion for cosmology. The methodology was extended to two scalar fields models and to deformed theories, as in references \cite{bglm,blp}. Another point to be emphasized is that such a formalism is well established in the standard Friedmann-Robertson-Walker (FRW) cosmology, and also in tachyonic scalar field dynamics, as pointed out in \cite{blrr,bglm,ablop,blr}. Moreover, a recent study shows some interesting aspects about twinlike behavior between the standard and the tachyonic dynamics \cite{bd}.

A great advantage of this method is that we can usually obtain analytical physical parameters in flat or curved space-time scenarios. However, when we deal with two real scalar fields systems, the first-order differential equations can be coupled nontrivially and, in several situations, they do not have analytical solutions. Because of this property the analytical models are limited to a few solvable examples when there are two real scalar fields. Thus, in order to study more general scalar field models with analytical defects, we are going to apply the extension procedure developed by Bazeia, Losano, and Santos \cite{bls} in the standard FRW cosmology. 

This extension map is based on the determination of two scalar fields models from the coupling between two single scalar field systems. As mentioned in reference \cite{bls}, a relevant feature of the extension method is that the analytical solutions of the one scalar field models satisfy the resultant two scalar fields system. In principle, by applying the method, we can infer the main behavior of the physical parameters since we are coupling two standard one-field models. However, this current procedure gives us unpredictable results for some physical parameters, which we discuss carefully.

\section{Generalities}
\label{sec:2}
In this section, we briefly review the first-order formalism considering one and two scalar fields coupled to gravity, as presented by Bazeia et al. in \cite{bglm}. Let us work with a two-field cosmological model described by the action
\be \label{eq1}
S=\int\,d^4\,x\,\sqrt{-g}\,\left[-\frac{R}{4}+{\cal L}(\phi_i\,,\p_\mu\,\phi_i)\right]\,.
\ee
In our notation, $i=1,2$, $\phi_{1}=\phi(t)$, $\phi_{\,2}=\chi(t)$ , $4\,\pi\,G=1$, $c=1$. Furthermore, here $g$ represents the determinant of the metric and $R$ is the Ricci scalar.  
The minimization of the action leads us to the equation of motion
\be
R_{\mu\nu}-\frac{1}{2}g_{\mu\nu}R=2\,T_{\mu\nu}\,,
\ee
with the following energy-momentum tensor:
\be
T_{\mu\,\nu}=2\frac{\p{\cal L}}{\p g^{\mu\nu}}-\,g_{\mu\nu}\,{\cal L}\,,
\ee
whose components are $T_{\mu\,\nu}=(\rho,-p,-p,-p)$, where $\rho$ and $p$ are the total energy density and pressure of the universe, respectively.

Once we are dealing with the standard FRW metric, the differential metric length has the form
\be
ds^2=dt^2-a^2(t)\left[\frac{dr^2}{1-kr^2}+r^2(d\theta^2+\sin^2\theta d\phi^2)\right]\,,
\ee
where $a(t)$ is the scale factor and $k$ is the curvature parameter. With $k=0$ we have a flat space-time, with $k=1$  spherical curvature, and with $k=-1$ hyperbolic geometry. Moreover, the Friedmann equations are 
\be 
H^2 =\frac{2}{3}\, \rho-\frac{k}{a^2}\,;\qquad H=\frac{\dot{a}}{a}\,,
\ee
and
\be
\frac{\ddot{a}}{a}=-\frac{1}{3}\left(\rho+3\,p\right)\,;\qquad \bar{q}=\frac{\ddot{a}\,a}{\dot{a}^{2}}=1+\frac{\dot{H}}{H^2}\,,
\ee
where $H$ and $\bar{q}$ are the Hubble and the acceleration parameters. Another relevant physical quantity is given by
\be \label{eq13}
\omega=\frac{p}{\rho}\,,
\ee
known as equation of state (EoS) parameter.

\subsection{First-Order Formalism for One Scalar Field in a Flat Space-Time}

We develop the first-order formalism considering only a single field, which means taking $\phi_{\,2}=0$ in $(\ref{eq1})$. Then the standard scalar Lagrangian is simply
\be
{\cal L}=\frac{\dot{\phi}^2}{2}-V(\phi)\,,
\ee
leading to the equation of motion
\be \label{eq2}
\ddot{\phi}+3\,H\,\dot{\phi}+V_{\phi}=0\,,
\ee
and in this description it is straightforward to check that 
\be
\rho=\dot{\phi}^2+V(\phi)\,;\qquad p=\dot{\phi}^2-V(\phi)\,
\ee
are the density and the pressure due to the scalar field. Furthermore, the Friedmann equations have the form
\be 
H^2=\frac{2}{3}\left(\frac{\dot{\phi}^{\,\,2}}{2}+V\right)-\frac{k}{a^2}\,; \qquad \dot{H}=-\dot{\phi}^{\,2}+\frac{k}{a^2}\,.
\ee
The next step is to define $H=W(\phi)$ and here we are interested in a flat space-time description, which means $k=0$, so  
\be
\dot{H}=W_\phi\,\dot{\phi}\,,
\ee
which directly leads us to the first-order differential equation for $\phi(t)$, given by
\be
\dot{\phi}=-W_\phi.
\ee
Moreover, the previous assumption for the Hubble parameter implies that the scalar potential has the form
\be
V=\frac{3}{2}\,W^{\,2}-\frac{W_\phi^{\,2}}{2}\,.
\ee
This general first-order formulation is also known as the Hamilton-Jacobi approach and more details can be found in Salopek and Bond \cite{salopek}, Kinney \cite{kinney}, and also in references \cite{markov,muslimov}.

As we mentioned, it is not easy to solve the cosmological equations of motion analytically, and sometimes it is necessary to use approximation methods as the {\textit{slow-roll}} regime \cite{linde,albrecht,liddle}. The {\textit{slow-roll}} approximation requires $\ddot{\phi}$ and $\dot{\phi}^{\,2}$ to be small, in such a way that the one-field dynamics are rewritten as
\be 
3\,H\,\dot{\phi}\simeq -V_\phi\,.
\ee
Moreover the assumption $\dot{\phi}^{\,2}/2 \approx V$ implies that
\be
H^{\,2}\simeq \frac{2}{3}\,V\,;\qquad \rho \approx V\,;\qquad \rho \approx -p\,,
\ee
meaning that $\omega \approx -1$. However, our approach is based on analytically solvable models. Consequently there is no sense in neglecting either $\ddot{\phi}$ or $\dot{\phi}^{\,2}$.

The same argument is valid for the effective two scalar field model. 


\subsection{First-Order Formalism for Two Scalar Fields in a Flat Space-Time}

In the two scalar fields approach, the Lagrangian density is
\be
{\cal L}=\frac{\dot{\phi}^{2}}{2}+\frac{\dot{\chi}^{2}}{2}-V(\phi,\chi)\,,
\ee
which leads us to the equations of motion 
\be
\ddot{\phi}+3\,H\,\dot{\phi}+V_{\phi}=0\,;\qquad \ddot{\chi}+3\,H\,\dot{\chi}+V_{\chi}=0\,.
\ee
Consequently, the energy density and pressure are
\be
\rho=\frac{\dot{\phi}^{2}}{2}+\frac{\dot{\chi}^{2}}{2}+V(\phi,\chi)\,;\qquad p=\frac{\dot{\phi}^{2}}{2}+\frac{\dot{\chi}^{2}}{2}-V(\phi,\chi)\,,
\ee
and we also determine the following expressions for the Hubble parameter in a flat space-time:
\be
H^2=\frac{\dot{\phi}^{2}}{3}+\frac{\dot{\chi}^{2}}{3}+\frac{2}{3}\,V(\phi,\chi)\,;\qquad \dot{H}=-\dot{\phi}^{2}-\dot{\chi}^2\,.
\ee
Then, by defining $H=W(\phi,\chi)$, we directly obtain the first-order differential equations
\be \label{eq3}
\dot{\phi}=-W_\phi(\phi,\chi)\,;\qquad \dot{\chi}=-W_\chi(\phi,\chi)\,,
\ee
and the scalar potential 
\be
V(\phi,\chi)=\frac{3}{2}\,W(\phi,\chi)^2-\frac{W_\phi(\phi,\chi)^2}{2}-\frac{W_\chi(\phi,\chi)^2}{2}\,.
\ee

This two-field description is also named ``hybrid inflation'' as pointed by Kinney in \cite{kinney}, where an approach based on the Hamilton-Jacobi formalism was applied. Kinney considered the scalar field matter equation of state as the fundamental quantity in the dynamical equations instead of the expansion rate. 

As mentioned, we search for analytic models, and a well-known technique to solve the expressions presented in $(\ref{eq3})$ is the integrating factor method rewriting the first-order differential equations as
\be \label{eq4}
\phi_\chi=\frac{d\,\phi}{d\,\chi}=\frac{W_\phi}{W_\chi}\,.
\ee
In general, this equation is nonlinear and its integration yields to a relation between the fields $\phi$ and $\chi$ known as the orbit equation.

These are the most important aspects of the first-order formalism used in this study. 

\section{The Extension Method}
\label{sec:3}

Here we summarize the basic theory behind the extension procedure by following the concepts presented in \cite{bls}. Let us begin by writing our field $\phi(t)$ as
\be
\phi=f(\chi)\,;\qquad \chi=f^{-1}(\phi)\,,
\ee
where the function $f(\chi)$ is invertible and called  the ``deformation function'' \cite{def_1}. We are also assuming that $\chi(t)$ describes another one-field theory model. The previous definition leads us to
\be
\dot{\phi}=f_{\chi}\,\dot{\chi}\,,
\ee
yielding the first-order differential equations
\be
\dot{\phi}=-W_\phi(\phi)=-f_{\chi}\,W_{\chi}(\chi)\,;\qquad \dot{\chi}=-W_\chi(\chi)\,.
\ee
The last relations can be rearranged as
\be \label{eq5}
\phi_\chi=f_\chi= \frac{W_\phi(\chi)}{W_\chi(\chi)}\,,
\ee
which has a structure similar to Eq. ($\ref{eq4}$). The main idea of the extension method is to use the deformation function and its inverse to express ($\ref{eq5}$) as
\begin{widetext}  
\be \label{eq6}
\phi_\chi=\frac{W_\phi}{W_\chi}\equiv\frac{a_1\,W_\phi(\chi)+a_2\,W_\phi(\phi,\chi)+a_3\,W_\phi(\phi)+c_1\,g(\chi)+c_2\,g(\phi,\chi)+c_3\,g(\phi)}{b_1\,W_\chi(\chi)+b_2\,W_\chi(\phi,\chi)+b_3\,W_\chi(\phi)}\,,
\ee
\end{widetext}
with $W_\phi(\phi)=W_\phi(\chi)=W_\phi(\phi,\chi)$, $W_\chi(\chi)=W_\chi(\phi)=W_\chi(\phi,\chi)$ and $g(\phi)=g(\chi)=g(\phi,\chi)$. Furthermore, the constraints $a_1+a_2+a_3=1$, $b_1+b_2+b_3=1$ and $c_1+c_2+c_3=0$ must be satisfied.

Thus we recognize $(\ref{eq6})$ as the first-order differential equation related to the orbit between the fields $\phi$ and $\chi$.

As is well known, the functions $W_\phi$ and $W_\chi$, in this effective system, need to obey 
\be \label{eq7}
W_{\phi\,\chi}=W_{\chi\,\phi}\,,
\ee
and from Eq. $(\ref{eq6})$ we can redefine $W_\phi$ as
\bn
W_\phi &=& a_1\,W_\phi(\chi)+a_2\,W_\phi(\phi,\chi)\\ \nonumber
&&
+a_3\,W_\phi(\phi)+c_1\,g(\chi)+c_2\,g(\phi,\chi)+c_3\,g(\phi)\,, \nonumber
\en
and $W_\chi$ as
\be
W_\chi=b_1\,W_\chi(\chi)+b_2\,W_\chi(\phi,\chi)+b_3\,W_\chi(\phi)\,.
\ee

Therefore, by applying $(\ref{eq7})$ we find the second constraint relation
\bn
&&
b_2\,W_{\chi\,\phi}(\phi,\chi)+b_3\,W_{\chi\,\phi}(\phi)=a_1\,W_{\phi\,\chi}(\chi) \\ \nonumber
&&
+a_2\,W_{\phi\,\chi}(\phi,\chi)+c_1\,g_\chi(\chi)+c_2\,g_\chi(\phi,\chi)\,,  \nonumber
\en
which we use to determine the function $g$, completing the necessary ingredients to calculate our effective superpotential for the two-field model.

\section{The Effective Model - Example}
\label{sec:4}
In this example, we use the extension procedure in order to construct an effective model with
\be
W(\phi)=A\,\phi^2+B\,;\qquad \dot{\phi}=-W_\phi(\phi)=-2\,A\,\phi\,,
\ee
with analytical solution
\be \label{eq8}
\phi(t)=e^{\,-2\,A\,t}\,,
\ee
and by 
\be
W(\chi)=\alpha\,\cosh(\beta\,\chi)\,; \,\,\, \dot{\chi}=-W_\chi(\chi)=-\alpha\,\beta\,\sinh(\beta\,\chi)\,,
\ee
where the field $\chi(t)$ is
\be \label{eq9}
\chi(t)=\frac{2}{\beta}\,\text{arccoth}\,\left(e^{\alpha\,\beta^2\,t}\right)\,.
\ee
These two models were studied in more detail in references \cite{bglm} and \cite{blp}. It is straightforward to check that the deformation function which connects the two systems is
\bn 
\phi&=&f(\chi)=\left\{\coth\left(\frac{\beta\,\chi}{2}\right)\right\}^{-\frac{2\,A}{\alpha\beta^2}}\,. \\ \nonumber
&& \nonumber
\en
We can write $W_\phi(\phi)$ and $W_\chi(\chi)$ in an equivalent form, using the deformation function. Such a procedure leads us to the set of equations
\bn
&&
W_\phi(\phi)=2\,A\,\phi\,, \\ \nonumber
&&
W_\phi(\chi)=2\,A\,\left\{\coth\left(\frac{\beta\,\chi}{2}\right)\right\}^{-\frac{2\,A}{\alpha\beta^2}}\,, \\ \nonumber
&&
W_\chi(\chi)=\alpha\,\beta\sinh(\beta\,\chi)\,, \\ \nonumber
&&
W_\chi(\phi)=2\,\alpha\,\beta\frac{\phi^{-\frac{\alpha\,\beta^2}{2A}}}{\phi^{-\frac{\alpha\,\beta^2}{A}}-1}\,,  \nonumber
\en
and for simplicity, we do not consider the forms $W_\phi(\phi,\chi)$ and $W_\chi(\phi,\chi)$, which is the same as taking $a_2=b_2=0$. We also choose $c_1=0$, implying  
\bn 
c_2\,g(\phi,\chi)&=&-2\,a_1\,A\left\{\coth\left(\frac{\beta\,\chi}{2}\right)\right\}^{-\frac{2\,A}{\alpha\beta^2}} \\ 
&&
+\frac{b_3\,\alpha^2\,\beta^3}{A}\,\frac{\phi^{\frac{\alpha\,\beta^2}{2A}-1}\left(\phi^{\frac{\alpha\,\beta^2}{A}}+1\right)}{\left(\phi^{\frac{\alpha\,\beta^2}{A}}-1\right)^2}\,\chi\,, \nonumber
\en
and by applying the deformation function, we find that
\bn\nonumber
c_2\,g(\phi)&=&-2\,a_1\,A\,\phi+\frac{2\,b_3\,\alpha^2\,\beta^2}{A}\,\frac{\phi^{\frac{\alpha\,\beta^2}{2A}-1}\left(\phi^{\frac{\alpha\,\beta^2}{A}}+1\right)}{\left(\phi^{\frac{\alpha\,\beta^2}{A}}-1\right)^2}\\ 
&&
\,\times\,\text{arccoth}\left(\phi^{-\frac{\alpha\,\beta^2}{2A}}\right)\,.
\en
With these ingredients, we determine that the superpotential for our effective two scalar field model is given by
\begin{widetext}
\be
W(\phi,\chi)=A\,\phi^2+B+2\,b_3\,\alpha\,\beta\frac{\phi^{-\frac{\alpha\,\beta^2}{2A}}}{\phi^{-\frac{\alpha\,\beta^2}{A}}-1}\,\chi 
-\frac{2\,b_3\,\alpha}{\phi^{\frac{\alpha\,\beta^2}{A}}-1}\,\left[1-2\,\phi^{\frac{\alpha\,\beta^2}{2A}}\text{arccoth}\,\left(\phi^{-\frac{\alpha\,\beta^2}{2A}}\right)\right]+b_1\,\alpha\,\cosh(\beta\,\chi)\,.
\ee

\end{widetext}

Therefore we can use the superpotential together with our analytical solutions (Eqs. $(\ref{eq8})$ and $(\ref{eq9})$) to compute the Hubble parameter $H(t)$, the scale factor $a(t)$, the acceleration parameter $\bar{q}(t)$, the EoS parameter $\omega(t)$, the density $\rho(t)$ and the pressure $p(t)$.

Here we focus on the simplest coupling configuration between the fields $\phi$ and $\chi$, corresponding to $b_1=1$ and $b_3=0$. The details concerning the behavior of the physical parameters for such a choice are shown in Figures \ref{FIG1}, \ref{FIG2}, and \ref{FIG3}. Furthermore, the explicit forms of $H(t)$, $a(t)$ and $\omega (t)$ are presented below: 
\begin{widetext}
\be \label{eq10}
H(t)=B+A e^{\,-4\, A\, t}+\alpha \, \cosh\left[2 \,\text{arccoth}\,\left(e^{\, \alpha \, \beta ^2\,t}\right)\right]\,,
\ee
\be \label{eq11}
a(t)=\,a_0\,\left[2\, \left(1-e^{\,2\, \alpha  \,\beta ^2\,t}\right)\right]^{\beta^{\,-2}}\,\exp\,\left[-\frac{1}{4} e^{-4\, A\, t}+\left(B-\alpha\right)\, t\right]\,,
\ee
\be \label{eq12}
\omega(t)=\frac{8\, A^2\, \cosh(4\, A \,t)-3 \left[B+A e^{-4\,A\,t}+\alpha\,\coth\left(\alpha \, \beta ^2\,t\right)\right]^2+2 \alpha ^2 \beta ^2 \text{csch}^2\,\left( \alpha \, \beta ^2\,t\right)-8\, A^2 \,\sinh(4\, A \,t)}{3 \left[B+A\, e^{-4 \,A\, t}+\alpha  \coth\left(\alpha \, \beta ^2\,t\right)\right]^2}\,.
\ee
\end{widetext}
We also obtain such analytical parameters in the case $b_3\neq 0$, which is shown in Fig.(\ref{FIG4}), where we plot the time evolution of $\omega$.

A remarkable feature of this effective hybrid model is that we can explore the cosmological parameters by means of the time evolution of the fields. Therefore we do not need to consider any kind of specific regime for $\phi(t)$ or $\chi(t)$, representing a more general description than those reported on previous studies concerning the two-field approach, as the one presented in \cite{kinney}.

The properties of the analytical parameters obtained above as well as their time evolution are discussed in more detail in the next sections.

\section{Dimensional Analysis}
\label{sec:5}

Here we want to motivate the cosmological interpretations reported later in Sec. \ref{sec:6}.

Firstly, we see from Eq.($\ref{eq11}$) that the non-dimensional property of the scale factor is respected, since it is given by the product of an exponential with an arbitrary non-dimensional constant. Recall that the scale factor, which is equal to $1$ at present and is independent of location or direction in FRW cosmology, tells us how the expansion of the universe depends on time.

The dimension of the Hubble parameter $H(t)$ in Eq. ($\ref{eq10}$) is directly connected to the dimension of the constants $A$, $B$ and $\alpha$. From ($\ref{eq10}$) we see that it would be interesting if those constants had the dimension inverse time, which is in fact the Hubble parameter dimension, since from Hubble's law, $v=H(t)r$, with $v$ being the recession (or approximation, in the case of the Local Group) velocity of the galaxy and $r$ the distance to it. Eq. ($\ref{eq11}$) only strengthens this assumption. One can see that for the argument of the first exponential to be dimensionless, $[\alpha]=[\beta]=[t]^{-1}$, as also required in the second exponential.

To solve the Friedmann equations for the energy density $\rho$ and pressure $p$, an EoS, i.e., a mathematical relation between $p$ and $\rho$, might be useful. For cosmological purposes, the EoS can be written in a linear form as Eq. ($\ref{eq13}$), with $\omega$ being a dimensionless number if we take $c=1$, since $[p]/[\rho]=[c]^2$. One can check that $\omega$ is, indeed, dimensionless in Eq. ($\ref{eq12}$).

\section{Cosmological Interpretations}
\label{sec:6}

\begin{figure}[ht!]
\vspace{1cm}
\includegraphics[{height=04cm,angle=00}]{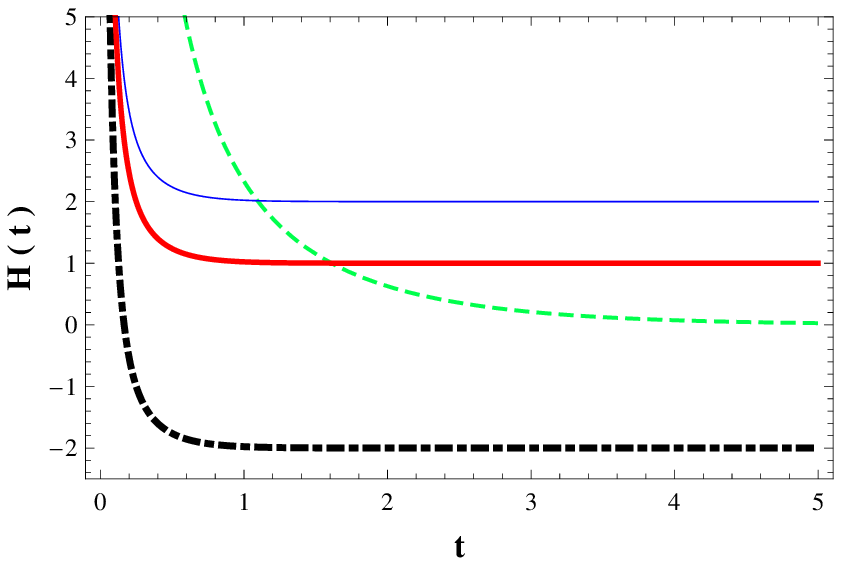}
\includegraphics[{height=04cm,angle=00}]{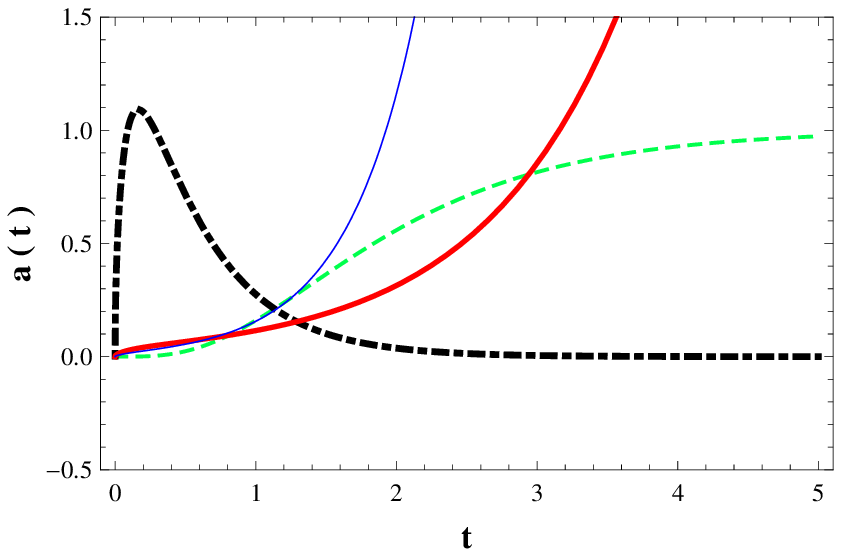}
\vspace{0.3cm}
\caption{Plots of parameters $H(t)$ and $a(t)$, where $A=5$, $\alpha=-1$, $\beta=3/2$, $a_0=3/2$ and $B=-3$ for the black (dot-dashed) curve, $a_0=1/32$ and $B=0$ for the red (thicker) curve, $a_0=1/64$ and $B=1$ for the blue (thin) curve, and $A=5$, $\alpha=-2$, $\beta=1/2$, $a_0=1/16$, $B=-2$ for the green (dashed) curve. The values of $a_0$ were chosen in order to show the parametric behavior for the different scenarios. Moreover, $a=1$ indicates the present value of the parameter.}
\label{FIG1}
\end{figure}

\begin{figure}[ht!]
\vspace{1cm}
\includegraphics[{height=04cm,angle=00}]{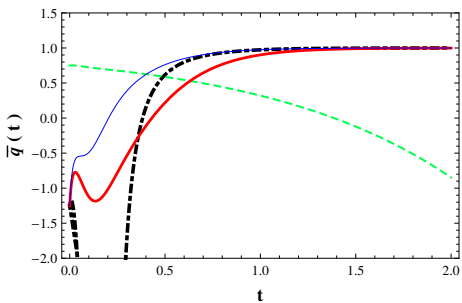}
\vspace{0.3cm}
\caption{Here we show the different forms of the acceleration parameter $\bar{q}(t)$, where we considered $A=5$, $\alpha=-1$, $\beta=3/2$ with $B=-3$ for the black (dot-dashed) curve, $B=0$ for the red (thicker) curve, $B=1$ for the blue (thin) curve, and $A=5$, $\alpha=-2$, $\beta=1/2$, $B=-2$ for the green (dashed) curve.}
\label{FIG2}
\end{figure}

\begin{figure}[h!]
\vspace{1cm}
\includegraphics[{height=04cm,angle=00}]{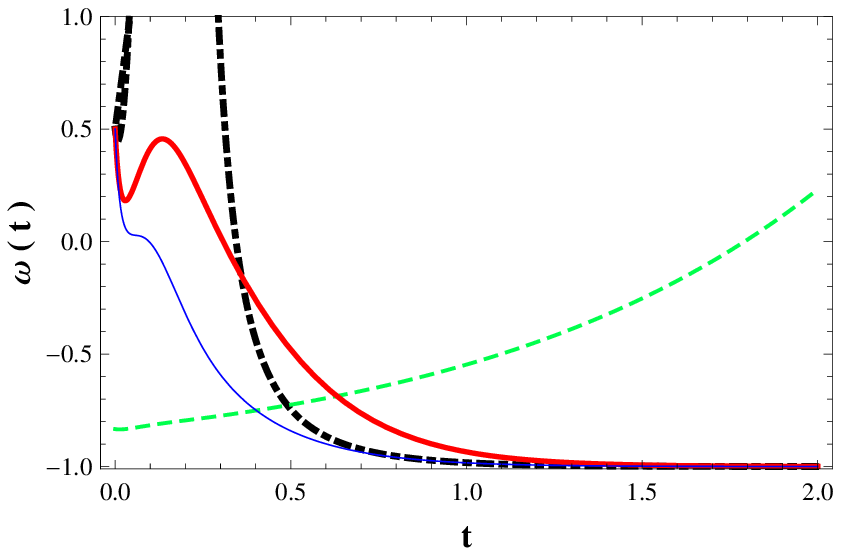}
\includegraphics[{height=04cm,angle=00}]{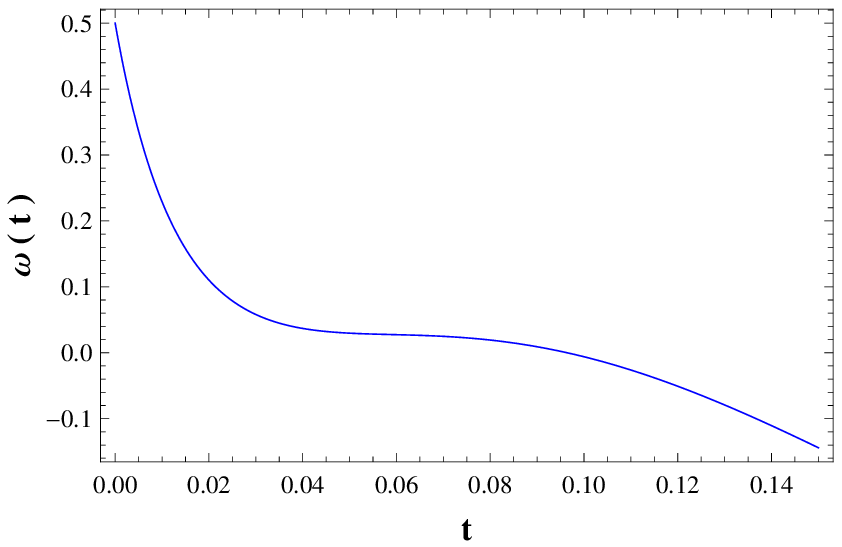}
\vspace{0.3cm}
\caption{Plots of $\omega(t)$, with  $A=5$, $\alpha=-1$, $\beta=3/2$, $B=-3$ in the black (dot-dashed) curve, $B=0$ in the red (thicker) curve, $B=1$ in the blue (thin) curve, and $A=5$, $\alpha=-2$, $\beta=1/2$, $B=-2$ in the green (dashed) curve. The figure in the lower panel shows in more detail the plateau-like behavior of $\omega(t)$, which occurs in the blue (thin) curve.}
\label{FIG3}
\end{figure}

\begin{figure}[h!]
\vspace{1cm}
\includegraphics[{height=04cm,angle=00}]{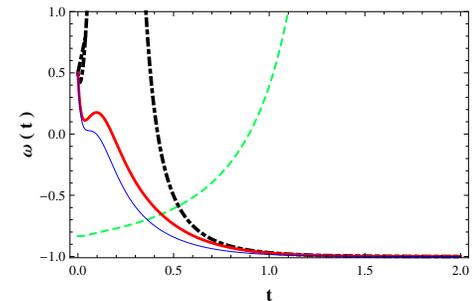}
\vspace{0.3cm}
\caption{This figure shows the EoS parameter with $b_3=b_1=1/2$, $A=5$, $\alpha=1$, $\beta=3/2$ and $B=-3$ for the black (dot-dashed) curve, $B=0$ for the red (thicker) curve and $B=1/2$ for the blue (thin) curve. We also present $A=5$, $\alpha=-2$, $\beta=1/2$ and $B=-2$ in the green (dashed) curve. Note the similarity between these results and those illustrated in the lower panel of Fig.(\ref{FIG3}).}
\label{FIG4}
\end{figure}

In this section we show that our model presents physical and cosmological consistence for some given values of $A$, $B$, $\alpha$ and $\beta$. The goal is to analyze Figs.(\ref{FIG1}-\ref{FIG4}) from the perspective of the cosmological parameters behavior predicted by the $\Lambda$CDM cosmological scenario.

Since $H\sim t^{-1}$, with $t$ being the Hubble time, $H(t)$ must decrease with time \cite{ryden/2003}, as observed in Fig.(\ref{FIG1}). Also, we discard the black (dot-dashed) curve once it allows negative values of $H(t)$, which is a physical inconsistency in an expanding universe, since from Eq.(5), $H(t)=\dot{a}/a$, where $a$ as a function of the redshift $z$ is given by $a=1/(1+z)$, in such a manner that it must increase as time passes by (redshift decreases).

An interesting feature about the black (dot-dashed) curve for $a(t)$ in Fig.(\ref{FIG1}) is the bump for small values of $t$. In the inflationary phase, when the energy density of the universe is dominated by a (cosmological) constant, the Friedmann equation solution is a scale factor that grows exponentially with time as $a(t)\propto e^{H_\iota t}$, with $H_\iota$ being the value of the Hubble parameter during inflation \cite{ryden/2003}. This bump might thus represent the inflationary phase. Nevertheless, in the present case, the black (dot-dashed) curve for $H(t)$ has been discarded, so for cosmological purposes, all the curves with $A=5$, $\alpha=-1$, $\beta=1.5$, $a_0=3/2$ and $B=-3$ must also be discarded.

\pagebreak
\noindent

However in Fig.(\ref{FIG1}), one can see that the green (dashed) curve for $a(t)$ represents $\dot{a}\rightarrow 0$ for large values of $t$, which implies a null Hubble parameter. This observation combined with the anomalous behavior of the acceleration parameter of the green (dashed) curve (see Fig.(\ref{FIG2})) configures an unpleasant cosmological scenario. Therefore, we focus our attention to the blue and red curves.

In Figs.(\ref{FIG3}-\ref{FIG4}) we plot the EoS parameter $\omega$. We zoom in on the blue (thin) curve in Fig.(\ref{FIG3}), and in order to clarify its features, let us briefly review some aspects concerning the density of the universe and the EoS parameter. 

The conservation of the energy-momentum tensor ($\nabla_\mu T^{\mu\nu}=0$) in standard Einstein's field equations results in
\be
\rho=\rho_0a^{-3(1+\omega)}\,
\ee
for the density of the universe if we consider $\rho_0$ a constant and $a_0=1$ the present value of the scale factor.
In the cosmology derived from general relativity, there are three regimes in which the universe dynamics is dominated respectively by radiation, matter and cosmological constant \cite{dodelson/2003}: the relativistic matter scenario, related to $\omega=1/3$ (which implies $\rho_r\propto a^{-4}$); the non-relativistic matter scenario, related to $\omega=0$ ($\rho_m\propto a^{-3}$); and the quantum vacuum scenario, corresponding to $\omega=-1$ ($\rho_\Lambda=\rho_0$).

From the blue (thin) curve in Fig.(\ref{FIG3}), note that for early times, $\omega$ assumes the value $1/3$ and values near to it, which shall represent the radiation dominated era. As the universe expands and cools down, the matter-radiation decoupling makes the universe propitious to form the stars and larger structures, as galaxies and clusters of galaxies. This era is dominated, then, by matter, with $p=0$ ($\omega=0$), which in Fig.(\ref{FIG3}) is presented as a plateau-like behavior of the blue (thin) curve for a non-negligible period of time. Note also that for high values of time, $\omega\rightarrow -1$, in agreement with recent observations of Planck satellite \cite{hinshaw/2013}, which by using Baryon Acoustic Oscillations and Cosmic Microwave Background data, have constrained the EoS parameter to $\omega=-1.073^{+0.090}_{-0.089}$.

\section{Final Remarks}
\label{sec:7}

Nowadays the first-order formalism based on one scalar field models is commonly used to describe quintessence scenarios for the accelerated expansion of the universe. Reference \cite{blp}, for instance, showed how to apply the deformation procedure in order to determine new analytical solutions for the one-field systems. In the two scalar fields description, there are several difficulties in integrating the dynamical equations. Furthermore, the standard approach of the deformation method is nontrivial to be implemented in this context. To search for new solvable models involving two scalar fields, we worked with the extension method, by coupling two single field models already studied in the literature.

The extension method applied to the two scalar fields formalism led us to plot Figs.(\ref{FIG1} - \ref{FIG4}). Some of the curves are excluded since they present a behavior that diverges from what is predicted by $\Lambda$CDM model. However, some of the plotted results, as the blue (thin) curves, have showed very interesting features which we revisit in the following. In Fig.(\ref{FIG3}) there is a plateau-like behavior around $\omega=0$ (consequently $p=0$) which could represent the matter-dominating era of the universe. The derivatives of $\omega$ with respect to time are near zero in the interval $t\in[0.04-0.10]$. Also, for $t<0.04$, one can see an abrupt variation of $\omega$ in a small interval of time. Note that this variation is continuous and constrained to values around $1/3$, which is the value of $\omega$ for a radiation-like EoS. Furthermore, the model predicts the late acceleration of the universe expansion since $\omega\to -1$ for high values of time.

We were also able to reproduce all the features related to the physical parameters expected by the one-field analysis, as we observed in reference \cite{blp}. In conclusion, the previous results support the two coupled scalar models description since new nontrivial behavior from the coupling between the fields was unveiled. Moreover, this extension method appears as a nice mathematical tool, which can be helpful when dealing with more complex quintessence models and even with a three-field coupling. Furthermore, it is also possible to implement this methodology in tachyonic dynamics and in scenarios with dust. Some of these applications are under investigation and we hope to report on them in the near future.

\acknowledgments

P. H. R. S. M. would like to thank CAPES and J. R. L. S. would like to thank CAPES and CNPq for financial support. The authors are grateful to Antonio Soares de Castro for reading the manuscript, offering suggestions for improvements, and also would like to thank the anonymous referee for the valuable comments and for bringing the references \cite{salopek,kinney} to their attention.


\end{document}